\documentstyle[12pt]{article}
\begin{document}

\textheight 580pt

\begin{titlepage}

\hfill{SISSA - REF 98/EP/142}

\vspace{1cm}

\centerline{\huge{Interacting Strings in Matrix String Theory}}

\vspace{1cm}

\centerline{\large{Giulio Bonelli}}

\vspace{1cm}

\centerline{International School for Advanced Studies and INFN}

\centerline{Trieste -- Italy}

\centerline{e-mail: bonelli@sissa.it}

\vspace{5cm}

{\bf Abstract:}
It is here explained how the Green-Schwarz superstring 
theory arises from Matrix String Theory. This is obtained as the strong 
YM-coupling limit of the theory expanded around its BPS instantonic
configurations, via the identification of the interacting string diagram 
with the spectral curve of the relevant configuration. 
Both the GS action and the perturbative weight $g_s^{-\chi}$,
where $\chi$ is the Euler characteristic of the world-sheet surface and
$g_s$ the string coupling, are obtained. 

\vspace{2cm}

To be published in the Proceedings of
{\it Quantum Aspects of Gauge Theories, Supersymmetry and Unification},
Corfu, Greece -- September 1998

\end{titlepage}
\def\ln{{\rm ln}}
\def\Az{{A_z}}
\def\Ab{{A_{\bar z}}}
\def\Dz{D_\z}
\def\Db{D_{\bar z}}
\def\bX{\overline X}
\def\A{{\cal A}}
\def\F{{\cal F}}
\def\G{{\cal G}_{w\bar w}}
\def\Z{{\bf Z}}
\def\d{\partial}
\def\dw{\partial_w}
\def\dz{\partial_z}
\def\dc{\partial_\zeta}
\def\dbw{\partial_{\bar w}}
\def\dbz{\partial_{\bar z}}
\def\dbc{\partial_{\bar \zeta}}
\def\sinh{{\rm sinh}}
\def\det{{\rm Det}}
\def\Tr{{\rm Tr}}
\newcommand\0{\nonumber}
\newcommand\ee{\end{eqnarray}}	 	
\newcommand\be{\begin{eqnarray}}
\def\l{\label}
\newcommand\ba{\begin{array}}			
\newcommand\ea{\end{array}}
\newcommand\bma{\left(\ba{cccccccc}}
\newcommand\ema{\ea\right)}			
\newcommand\e{{\rm e}}
\def\o{{\omega}}
\newcommand\cH{{\cal H}}
\newcommand\cP{{\cal P}}
\newcommand\Lm{{\Lambda}}

\section{Introduction}
During the last few years it has been developed 
a new approach to string theory.
This new approach arose to embody the dream of a non perturbative formulation
of string theory. This theory has been called M-theory \cite{M}, 
where however M, by now, still stands for Moon. In fact,
notwithstanding the great activity in this direction,
the full structure of M-theory remains largely elusive.
The major proposal for its definition has been given in \cite{bfss}
where it has been conjectured to be a quantum mechanical
system of $9+1$ bosonic matrices plus a fermionic counterpart which
carries the model to have a dimension 16 (nonlinear) supersymmetry.
In this formulation the dimension of the matrices is to be taken to infinity
to generate extended objects (M-branes) and the resulting theory
should describe M-theory in the infinite momentum frame.
A possible way to confirm this hypothesis is to understand 
if it does really reproduce, in the appropriate corners of the moduli space,
the known features of perturbative string theories.
The matrix string theory program \cite{motl,dvv}
\footnote{Let me thank T. Banks and L. Motl for pointing me out reference 
\cite{motl} which was lacking in the previous version of this proceeding
report.} has been formulated
in the conjectured neighborhood of the type IIA superstring theory.
Here the theory is realized as an $U(N)$ SYM with (8,8) supersymmetry
on a cylinder and the thesis is that its strong coupling limit
should describe type IIA with $g_s\sim g_{YM}^{-1}$.

For what concerns the free theory reproduction, things are not so difficult
to realize. The situation is as follows: consider the unique limit 
$$
\left(\matrix{ U(N) \cr SYM\cr {\cal N}=(8,8)\cr}\right)
\qquad
\buildrel {g_{YM}}\to\infty \over \longrightarrow 
\qquad
\left(\matrix{ \left({\bf R}^8\right)^N/S_N \cr CFT\cr {\cal N}=(8,8)\cr
}\right)
$$
Due to the fact that there is no interacting realization of 
the $D=2$, $(8,8)$-superconformal algebra, 
the IR CFT is forced to be the free theory 
twisted by the $U(N)$-Weyl group. Moreover,
the orbifold sectors of the theory, which are
identified with $S_N$ classes 
$[g]=(1)^{n_1}\cdot (2)^{n_2}\cdot\dots\cdot (N)^{n_N}$,
where $N=\sum_{a=1}^N an_a$, get a natural string interpretation as
states composed of $\sum_a a$ free strings each of length $n_a$.
This length is then identified with the discretized light-cone momentum
in appropriate units.
In \cite{dvv} there is also a step forward in the direction of understanding
the interacting perturbative string regime. 
The starting point is the observation that
in the CFT the string states (the orbifold sectors) 
are orthogonal. Therefore, to let strings interact, one should exit the 
conformal point with some vertex.
The conjectured DVV vertex is essentially the 
Mandelstam string vertex \cite{mandelstam} and properly generates the 
superstring perturbative expansion in the light-cone (see 
also R. Dijkgraaf's lecture in this volume). 

The problem we want to tackle here is how and where to find superstring 
interaction in the very structure of the SYM theory.
We will start with 
looking for (punctured) Riemann Surfaces
in the theory: if they would be found, then these surfaces 
should be identified
with the interpolating world-sheet between different string states.
Before entering in any detail, let us try to give some reasonable
form in which interpolating surfaces could appear.
As we saw above, the relevant string theory should arise already 
in the light-cone gauge. Therefore, one is led to look for the relative
Mandelstam diagrams which are, as it has been fully explained
in \cite{giwo}, representable as branched coverings of a cylinder. 
Let us first
review very quickly, as a preliminary point, the basics of how Riemann 
curves can be represented as branched coverings.

Let $z$ be a coordinate on a connected set ${\cal A}\subset {\bf CP^1}$
and let $a_i(z)$, $i=0,\dots,N-1$, be analytic functions on ${\cal A}$.
Let also $x\in{\bf CP^1}$ be an indeterminate variable.
Consider the curve $\Sigma$ in ${\cal A}\times{\bf CP^1}$ defined by
the polynomial equation
$$
P(x)=x^N+\sum_{i=0}^{N-1}a_i(z)x^i=
\prod_{k=1}^N\left(x-x_k(z)\right)=0
$$
and notice that, for generic $a_i$, the root functions $x_k(z)$
are not one-valued functions on ${\cal A}$. In fact they can exchange 
by continuing along paths encircling points where two (or more)
of them coincide. These points are called the branching points 
of the covering.
The covering structure of $\Sigma$ is given in terms of the copies
${\cal A}_k={\rm Im}\left({\cal A},x_k\right)$
in the following way: on each copy coherently give a cuts system
connecting the branching points and possibly the boundary 
$\partial{\cal A}_k$, 
then glue them together along the cuts in the way dictated
by the exchange in the roots set to get the surface.
In the following, ${\cal A}$ will be taken to be ${\bf C}-\{0\}$, 
which is an infinite
cylinder ${\cal C}$.

The following results have been obtained in collaboration with
L. Bonora and F. Nesti \cite{bbn1,bbn2}.
\section
{(4,4) preserving instantons and the strong coupling
 expansion of the partition function around them}
What we are going to show in the rest of the talk is how the above 
program can be realized 
if one looks at the instantonic sector of the theory.
More precisely, we will find a rich mathematical structure in that sector
and perform a full stringy interpretation of the strong coupling expansion 
of the theory around the generic instantonic configuration.
The outcome will be the Green -- Schwarz IIA superstring partition function.
\subsection{(4,4) instantons}
As a first step, let us explain the emergence of the relevant 
Riemann surfaces from the instanton equations.
The bulk action of the theory is
$$ 
S=\frac{1}{\pi} \int_{\cal C} d^2w \,\Tr \left(
D_w X^i D_{\bar w} X^i - \frac{1}{4g^2} F_{w\bar w}^2 -
\frac{g^2}{2}[X^i,X^j]^2+\right.
$$ $$\left.
+i (\theta_s D_{\bar w} \theta_s + \theta_c
D_w \theta_c) + 2ig \theta_s \gamma_i [X^i,\theta_c] \right)
$$
where $(X^i,\theta_s,\theta_c)$s are in the adjoint w.r.t. the $U(N)$ 
gauge group
and in the $(8_v, 8_s, 8_c)$ of the $SO(8)$ R-symmetry group respectively.
The gauge connection is of course an R-singlet.
This action is invariant under the following 
{\bf ${\cal N}=(8,8)$-supersymmetry} transformations
$$
\delta X^i = \frac{i}{g} (\epsilon_s \gamma^i \theta_c +
\epsilon_c \tilde\gamma^i \theta_s)
\, ,\quad\quad
\delta A_w = -2\epsilon_s \theta_s, \quad\quad \delta A_{\bar w}=
-2\epsilon_c \theta_c
$$
$$
\delta \theta_s = (-\frac{i}{2g^2} F_{w\bar w}  +
\frac {1}{2} [X^i,X^j]\gamma_{ij}) \epsilon_s -\frac{1}{g}D_w X^i
\gamma_i\epsilon_c
$$ $$
\delta \theta_c = (\frac{i}{2g^2} F_{w\bar w} +
\frac {1}{2} [X^i,X^j]\tilde\gamma_{ij}) \epsilon_c 
-\frac{1}{g}  D_{\bar w} X^i \tilde\gamma_i\epsilon_s
$$ 
There exists a full class of 
$(4,4)$-susy preserving classical configurations \cite{give,bbn1}:
$\theta_s=0$, $\theta_c=0$,
$X^i=0$ for $i=3,\dots,8$
while $X=X^1+iX^2$ and the connection $A$ satisfy the 
Hitchin system \cite{hitchin}
\be
F_{w\bar w}+ig^2\left[X,\bar X\right]=0\, ,
\quad
D_wX=0\, ,
\quad
D_{\bar w}\bar X=0\, .
\l{hitchin}\ee
This system is known to be integrable in terms of spectral curves.
These spectral curves are what we were looking for. To see them explicitly,
parametrize with full generality the 
above fields as
\be
X = Y^{-1}M Y \quad {\rm and}\quad A_w = -i Y^{-1}\d_w Y\,\,\, , 
Y\in SL(N,{\bf C})\l{para}\ee
where $Y$ and $M$ are still well defined fields on the cylinder.

As for $M$, it satisfies $\dw M=0$ without any further restriction.
Consider now the polynomial
$$P_X(x)=\det (x - X)=\det (x -M)=
x^N+\sum_{i=0}^{N-1}x^ia_i,$$
where $x$ is a complex indeterminate.
Since $\partial_w M=0$, we have $\dw a_i=0$ which means that 
the set of functions $\{a_i\}$ are antianalytic on the cylinder.
Therefore the equation 
\be P_X(x)=0 \l{P=0}\ee
identifies in the
$(w,x)$ space a Riemann surface $\Sigma$, which is an N--sheeted branched 
covering of the cylinder ${\cal C}$.
We can choose $M$ to be in a standard form as
$$
M=\left(\matrix{-a_{N-1}& -a_{N-2}& \ldots & \ldots  & -a_0\cr
		  1      & 0       & \ldots & \ldots  & 0  \cr
		  0       & 1       &  0  & \ldots  & 0 \cr
		  \ldots	   & \ldots     &  \ldots& \ldots  & 
		  \ldots \cr		
                  0	    & 0       & \ldots & 1    & 0 \cr
}\right)
$$
Notice that the branched covering structure is completely encoded in $M$ 
and is independent on the value of the coupling. 
This is the rigid part of the (4,4)-preserving instanton under the coupling 
flow.

As for $Y$, it contains all the dependence on the value of the coupling 
and is determined by the following deformed WZNW equation
\be
\dw\left(\dbw\Omega\Omega^{-1}\right)+g^2\left[M,\Omega 
M^+\Omega^{-1} \right]=0\,\, ,
{\rm where}\,\, \Omega\equiv YY^+\,\, .
\l{defo}\ee

The calculation we are going to perform is a strong coupling expansion 
of the partition function. As a necessary step, we need to
understand what the fate is of our instantons in this strong coupling limit
\cite{bbn2}.

As it is immediate from the Hitchin system equations (\ref{hitchin}),
at strong coupling we have
$$
\left[X,\bar X\right]=0 \quad \Rightarrow \quad
\matrix{
 X=U \hat X U^+\,\,\, , U\in U(N) 
\cr
\hat X={\rm diag}\left( x_1,\ldots , x_N \right)
\cr}
$$
where $x_i$ are the roots of (\ref{P=0}).
On the other hand we parametrized the solutions as
(\ref{para}), so we have
$$
X=Y_s^{-1} M Y_s\,\, , {\rm where} \,\, Y_s \,\, {\rm is }\,\, Y
{\rm at}\,\, g\sim\infty
$$
Therefore,
diagonalizing $M=S\hat X S^{-1}$ with $S_{ij}\equiv (x_j)^{N-i}$, we get
$Y_s=SU^+$ and $\Omega_s=Y_sY_s^+=SS^+$ satisfies, coherently with
(\ref{defo}),
$\,\,\dw\left(\dbw \Omega_s \Omega_s^{-1}\right)=0$.

Summarizing, at strong coupling the instantonic configuration is
\be
A_w=-i U\dw U^+ \,\, {\rm and}\,\, X=U \hat X U^+  
\l{back}\ee
Notice that along fixed time curves on the cylinder ${\rm Re}\, w
=T$ we get
$$
\hat X\,\,\,\rightarrow P_T^+\cdot \hat X \cdot P_T \quad {\rm and}\quad
U\,\,\,\rightarrow U\cdot P_T$$
with $P_T\in S_N$
describing the intermediate string state at time $T$ in the way described 
at the beginning of the talk.
Moreover, the unitary field $U$ defines a Cartan subalgebra 
${\tt t}= U{\tt t}_d U^+$, where ${\tt t}_d$ is the diagonal one.
This will be the Cartan subalgebra we will choose to split the 
fields in the strong coupling limit expansion of the theory.
\subsection{Expanding the action functional}
We are now going and face the problem of performing the strong 
coupling expansion of the action. Let us write the bulk action of 
the theory as
\footnote{Notice that the following holds up to boundary terms.
These terms are inessential for determining the bulk
structure of the theory at strong coupling. Nevertheless, they could 
become very interesting once one would like to have a full control of 
the theory at the boundary of the cylinder. In such a refined analysis, 
one should start from the beginning with a boundary control on the form of 
the action to start with.}
$$
S=\frac{1}{\pi} \int_{\cal C} d^2w \,\Tr\left(
D_w X^I D_{\bar w} X^I 
-\frac{g^2}{2}[X^I,X^J]^2
-g^2[X^I,X][X^I,\bX]+
\right.
$$ $$
\left.
+D_w X D_{\bar w} \bX 
-\frac{1}{4g^2}\left( F_{w\bar w} +i g^2[X,\bX]\right)^2
+i (\theta^-_s D_{\bar w} \theta^-_s + \theta^+_c
D_w \theta^+_c) + ig \theta^T \Gamma_i [X^i,\theta] \right)
$$ 
where $I= 3,4,...,8$.
To perform the expansion around any given instanton, 
write any field $\Phi$ as
$$
\Phi=\Phi^{(b)}+\phi^{\tt t}+\phi^{\tt n}\equiv 
\Phi^{(b)}+\phi\equiv \Phi^\circ +\phi^{\tt n}\,,
$$
where
$\Phi^{(b)}$ is the background value of the field (which is (\ref{back})), 
$\phi^{\tt t}$ are the fluctuations along the Cartan 
directions and $\phi^{\tt n}$ are the fluctuations 
along the non -- Cartan directions.

It is appropriate at this point to 
fix the gauge of the theory in the following way
$$
\G=D^\circ_w a_{\bar w} +D^\circ_{\bar w} a_w + i g^2 ([X^\circ, \bar x]
+ [{\bar X}^\circ, x])+ 2i g^2 [ X^{\circ I}, x^I]  =0\,,
$$
and to apply the Faddeev--Popov procedure by adding to the action
$$
S_{FP}=
S_{gf}+S_{ghost}=\frac{1}{4\pi g^2}\int d^2w~\G^2
 -\frac{1}{2\pi g^2} \int d^2w ~\bar c \frac {\delta \G}{\delta c} c\,,
$$
where $\delta$ represents the gauge transformation with parameter $c$.
We get a total action
$$
S_{tot.}=S+S_{FP}$$

To extract the leeding terms of the action, rescale fields as 
$$
A_w = A^{(b)}_w + g a_w^{\tt t} + a_w^{\tt n}, \quad
X= X^{(b)} + x^{\tt t} + \frac {1}{g} x^{\tt n},\quad
X^I = x^{I{\tt t}} + \frac{1}{g} x^{I{\tt n}},
$$ $$
\theta= \theta^{\tt t} +\frac{1}{\sqrt{g}} \theta^{\tt n}, \quad
c= g c^{\tt t} + \sqrt g  c^{\tt n}, \quad
\bar c= g \bar c^{\tt t} + \frac {1}{\sqrt g} \bar c^{\tt n}
$$
It is important to
notice
that these rescalings induce a unit Jacobian in the path integral measure
of the non--zero modes, but they may produce a non-trivial factor due to
the presence of zero modes.
After the above rescalings the action becomes
$$
S=S_{sc}+Q_{\tt n}+o\left(\frac{1}{\sqrt{g}}\right)\0\,,
$$
where
$$
S_{sc}=
\frac{1}{\pi} \int_{{\cal C}} d^2w \,\Tr \left[
D^{(b)}_w x^{I{\tt t}} D^{(b)}_{\bar w} x^{I{\tt t}} 
+ D^{(b)}_w x^{{\tt t}} D^{(b)}_{\bar w} \bar x^{{\tt t}} 
 +i (\theta^{\tt t}_s D^{(b)}_{\bar w} \theta^{\tt t}_s 
+ \theta^{\tt t}_c D^{(b)}_ w \theta^{\tt t}_c)\right.
$$ $$\left. +D^{(b)}_w a^{\tt t}_{\bar w}
D^{(b)}_{\bar w} a^{\tt t}_w + 
D^{(b)}_w \bar c^{{\tt t}} D^{(b)}_{\bar w} c^{{\tt t}}
\right]
$$
and $Q_{\tt n}$
is a quadratic term in $\phi^{\tt n}$.

Let us now show that the integration along non-Cartan directions does not 
contribute to the effective action. The exact expression for $Q_{\tt n}$
is 
$$
Q_{\tt n}= \frac {1}{\pi} \int d^2w \Tr \left[ \bar 
x^{\tt n}{\cal Q} x^{\tt n}+  
x^{I{\tt n}}{\cal Q} x^{I{\tt n}}+ 
a_{\bar w}^{\tt n}{\cal Q} a_w^{\tt n}+
\bar c^{\tt n}{\cal Q} c^{\tt n}+
i (\theta_s^{\tt n}, \theta_c^{\tt n}) {\cal A}
\left(\matrix{\theta_s^{\tt n}\cr \theta_c^{\tt n}\cr}\right) 
\right]\,,\quad {\rm where}
$$
$$
{\cal Q} = {\rm ad}_{X^{i\circ}}\cdot {\rm ad}_{X^{i\circ}}+
{\rm ad}_{a_{\bar w}^t}\cdot {\rm ad}_{a_w^t} 
\quad {\rm and}\quad
{\cal A} =\left( \matrix { i {\rm ad}_{a_{\bar w}^t} & \gamma_i 
{\rm ad}_{X^{\circ i}} \cr
\tilde \gamma_i {\rm ad}_{{\bar X}^{\circ i}} & 
i {\rm ad}_{a_{ w}^t}\cr}\right)\0\,.
$$
Notice that $Q_{\tt n}$ is a purely algebraic quadratic term in the 
$\phi^{\tt n}$ fluctuations which can be easily integrated over
without any zero-mode problem contribution to the path -- integral measure.
The integration over
$a^{\tt n}$ and  $c^{\tt n}$ exactly cancels to $1$ and also
the integration over $x^{\tt n}$ and $\theta^{\tt n}$
gives again $1$ due to supersymmetry
(${\cal A}{\cal A}^\dagger= {\cal A}^\dagger 
{\cal A}= - {\cal Q}$). 
Summing up the net result of integrating over the 
non--Cartan modes is $1$ and,
in the strong coupling limit, we are left with the
action $S_{sc}$ over the Cartan modes.
\subsection{Lifting the action to the world -- sheet}
Let us now show that $S_{sc}$
corresponds to the Green--Schwarz superstring action plus a free Maxwell
action on the world-sheet identified with the spectral curve of the 
relevant background instanton. The free Maxwell sector will be 
integrated out at the end of the story. The result of the integration 
along this sector will be a nice expected contribution.

Begin 
rewriting $S_{sc}$ in a diagonal 
representation of the background just undoing the $U$ rotation relative to 
the background structure (\ref{back}). The covariant derivative 
$D^{(b)}_w$ becomes the simple derivative $\dw$
and the Cartan subalgebra gets rotated to the diagonal one
$$
S_{sc}=
\frac{1}{\pi} \int_{{\cal C}} d^2w \,\Tr \left[
\d_w x^{I{\tt t}_d} \d_{\bar w} x^{I{\tt t}_d} 
+ \d_w x^{{\tt t}_d} \d_{\bar w} \bar x^{{\tt t}_d} 
 +i (\theta^{{\tt t}_d}_s \d_{\bar w} \theta^{{\tt t}_d}_s 
+ \theta^{{\tt t}_d}_c \d_ w \theta^{{\tt t}_d}_c)+\right.
$$ $$
\left. +\d_w a^{{\tt t}_d}_{\bar w}
\d_{\bar w} a^{{\tt t}_d}_w + 
\d_w \bar c^{{\tt t}_d} \d_{\bar w} c^{{\tt t}_d}
\right]
$$
Since all the matrices are diagonal we can rewrite this action in terms 
of the diagonal modes 
$\phi^{{\tt t}_d} = {\rm diag}\left(\phi_{(1)},\ldots,\phi_{(N)}\right)$
$$
S_{sc}=
\frac{1}{\pi} \int_{{\cal C}} d^2w \sum_{n=1}^N \Big[
\d_w x^i_{(n)} \d_{\bar w} x^i_{(n)}
+i (\theta_{s(n)} \d_{\bar w} \theta_{s(n)} 
+ \theta_{c(n)} \d_ w \theta_{c(n)})+
$$ $$
 +\d_w a_{\bar w(n)}
\d_{\bar w} a_{w(n)} + 
\d_w \bar c_{(n)} \d_{\bar w} c_{(n)}
\Big]
$$
As anticipated,
the individual components $\phi_{(i)}$ are not well defined 
fields on the cylinder and to give a meaning to the theory we must understand 
if they can be considered to be well defined fields on some other space.
To do this, observe that since 
$\phi^{{\tt t}}=U\phi^{{\tt t}_d}U^+$
is well-defined on the cylinder
and since following once ${\rm Re}\, w=T$, 
$
U\rightarrow U\cdot P_T$ with $P_T\in S_N$, then,
along ${\rm Re}\, w=T$, we get
$
\phi^{{\tt t}_d}
\rightarrow P_T^+ \cdot \phi^{{\tt t}_d}
\cdot P_T$.

What we want to show now is that these are exactly the properties 
a set of fields on a cylinder should have to be the representation
of a single local field on a Riemann surface
represented as a branched covering of the cylinder.

If $\Sigma$ is a branched covering of the cylinder ${\cal C}$, then there 
exists a projection map $\pi:\Sigma\,\,\to\,\,{\cal C}$
whose local inverse image is N-valued
$$
\pi^{-1}:\, w\,\,\to\,\, \left(x_{1}(w),\dots,x_{N}(w)\right)\, ,
$$
where $\{x_{i}(w)\}$ is the set of the roots of its polynomial 
equation (\ref{P=0}).
So, let $\tilde{\psi}$ be a local complex field on $\Sigma$:
$\pi_\star\tilde{\psi}=\left(\psi_{(1)}(w),\dots,\psi_{(N)}(w)\right)$
represents the field on each copy of the cylinder ${\cal C}$ 
composing the covering $\Sigma$ and
the $\psi_{(i)}(w)$'s are related exactly
by the $P_T$ monodromy along the curves ${\rm Re}\, w =T$. 

From the point of view of $\Sigma$, the $w$ coordinate is locally
defined via an abelian differential $\omega=dw$ with imaginary periods
\cite{giwo}.
This generates the factors needed to keep into 
account the differential weights of the various fields.

All this implies that
the field $\phi^{{\tt t}_d}$ represents a well-defined field on $\Sigma$
when rescaled with the appropriate $\omega$ factor
and we can lift the action to the Riemann surface $\Sigma$ obtaining
$$ 
S_{sc}= S^\Sigma_{GS} + S^\Sigma_{Maxwell}\, ,$$
where
$$
S^\Sigma_{GS}=\frac{1}{\pi} \int_{\Sigma} d^2z \left( 
\d_z \tilde x^i\d_{\bar z} \tilde x^i
+i (\tilde\theta_{s} \d_{\bar z} \tilde\theta_{s} 
+\tilde \theta_{c} \d_z \tilde\theta_c)\right)$$ 
$$
S^\Sigma_{Maxwell}= \frac{1}{\pi} \int_{\Sigma} d^2z\left(
g^{z\bar z}
\d_z \tilde a_{\bar z}
\d_{\bar z} \tilde a_{z} + 
\d_z \tilde {\bar c} \d_{\bar z}\tilde c\right)\,
$$
and the metric in the Maxwell term is 
$g_{z \bar{z}}=\omega_z\omega_{\bar z}$ with $z$ a system of local 
coordinates on $\Sigma$.

An expected
nice present from the Maxwell sector will be soon received.
To get it, let us integrate over this sector.
Since the action is quadratic
the integration produces a ratio of determinants, which turns out to be 
a constant (there is no dynamics for a massless vector field
in two dimensions), but we have to take account of the
zero modes for the fields that have been rescaled:
$$
\tilde a_z\,\to g\, \tilde a_z\, ,\quad \tilde a_{\bar z}\,\to g\, 
\tilde a_{\bar z}\, ,\quad
\tilde c\,\to g\,\tilde c\, ,\quad
\tilde {\bar c}\,\to g\, \tilde{\bar c}\, .
$$
The Maxwell partition function is then
$$
Z^{\Sigma}_{Maxwell}=
\int{\cal D}\left[\tilde{a},\tilde{c}\right]
\,\, e^{-S^{\Sigma}_{Maxwell}(\tilde{a},\tilde{c})}
=\frac{
{\det}'\nabla_c}{ {{\det}' \nabla_a}}
\propto g^{\P_c-\P_a}$$
where $\nabla$ 
denotes the relevant laplacian, $'$ means that 
the zero modes have been excluded from the computation 
of the regularized determinants
and $\P$ is the number of these zero modes.

As for the ghost fields, which are scalars, the only 
zero modes of the $\nabla_c$ operator on $\Sigma$ is the constant.
The zero modes of the Maxwell field correspond instead to the 
harmonic 1-differentials on $\Sigma$. 
If $\Sigma$ were a closed Riemann surface of genus $h$, their number 
would be $h$.
But $\Sigma$ is a Riemann surface with boundaries $b$ and the counting needs
a little trick to be performed.
Construct the double $\hat\Sigma$ of $\Sigma$
$\left(\hat\Sigma\sim\Sigma\times{\bf Z}_2\right)$:
$\hat \Sigma$ has genus $\hat h=2h +b-1$ and $\hat b=0$
and the number of analytic differential on $\Sigma$ 
that extend to $\hat\Sigma$ (analytic Schottky differentials) is 
$\hat h=2h +b-1$.
Summing up, we have therefore
$\P_c-\P_a=1-\hat h=2-2h-b=\chi_\Sigma$ and we get
$$
Z^{\Sigma}_{Maxwell}\quad\propto\quad
\left(\frac{1}{g}\right)^{-\chi_\Sigma}\, .$$
\section{The string theory interpretation}
Let us recollect the various terms to reconstruct the strong coupling 
limit of the $(8,8)$ YM partition function:
\be
Z_{sc}\quad
\huge{\sim}\quad
\int_{{\cal M}_{sc}} dm 
\left(1/g\right)^{-\chi}
\left[{\rm Jac}\right] 
\int D\left[\tilde x,\tilde\theta\right]
e^{-S_{GS}\left[\tilde x,\tilde\theta\right]}
\l{reco}\ee
${\cal M}_{sc}$ is the space of instantons at strong coupling: in this regime
each instantonic configuration is determined uniquely by a branched 
covering of the cylinder $\Sigma$, i.e. by a Mandelstam diagram;
$dm$ is the field theory induced measure on ${\cal M}_{sc}$: the integral in 
this sector is split as a sum over $h$ and an integral over 
${\cal M}_{sc}^h$;
in the $\left(1/g\right)^{-\chi}$ factor,
$\chi=2-2h-b$ is the Euler characteristic of $\Sigma$;
the $\left[{\rm Jac}\right]$-obian factor has been produced by
the background dependent field splitting we performed
(it depends of course on $\Sigma$) and
$S_{GS}$ is the Green-Schwarz superstring action on $\Sigma$

At this point, looking at (\ref{reco}), one is tempted to say
that MST in its strong coupling regime reproduces 
a discretized version of
the perturbative type IIA superstring theory 
in the light-cone with $g_s\propto 1/g$

Indeed we proved the above statement up to a couple of technical points:
prove that $dm\cdot{\rm [Jac]}$ generates the right 
superstring measure on the moduli space and 
complete the analysis of the theory at the boundary 
of the cylinder to reconstruct the boundary terms in the Mandelstam 
light-cone string.

Higher order terms in the expansion of the partition function 
should then represent non perturbative contributions to string 
theory.
One should also be able to 
include in the analysis D-branes. Up to D-particles,
this seems to be done with a careful sight 
at the $N\to\infty$ limit. Are there other relevant subleading saddle-points 
to consider to get the full theory? 

%
%

\end{document}